\documentclass[sigconf]{acmart}

\usepackage{enumitem}
\usepackage{colortbl}
\usepackage{bbm}
\usepackage{algorithm}
\usepackage{algorithmic}
\newcommand{\syndy}{\textsc{SynDy}}
\usepackage{float}
\newfloat{algorithm}{t}{lop}
\usepackage{amsmath}
\usepackage{verbatimbox}
\usepackage{tabularray}
\usepackage{array}
\newcolumntype{P}[1]{>{\centering\arraybackslash}p{#1}}
\usepackage[font={small},labelfont=bf]{caption}

\usepackage{microtype}

\newcommand{\inlinemath}[1]{{\begin{math}#1\end{math}}}

\newcommand{\riet}{{\sc Riet}}
\newcommand{\rietlab}{{\sc Riet Lab}}

\AtBeginDocument{%
  \providecommand\BibTeX{{%
    \normalfont B\kern-0.5em{\scshape i\kern-0.25em b}\kern-0.8em\TeX}}}

\setcopyright{acmcopyright}
\copyrightyear{2024} 
\acmYear{2024} 
\setcopyright{rightsretained} 
\acmConference[SIGIR '24]{Proceedings of the 47th International ACM SIGIR Conference on Research and Development in Information Retrieval}{July 14--18, 2024}{Washington, DC, USA}
\acmBooktitle{Proceedings of the 47th International ACM SIGIR Conference on Research and Development in Information Retrieval (SIGIR '24), July 14--18, 2024, Washington, DC, USA}
\acmDOI{10.1145/3626772.3657667}
\acmISBN{979-8-4007-0431-4/24/07}

\begin{document}

\title[\syndy{}: \underline{Syn}thetic \underline{Dy}namic Dataset Generation Framework for Misinformation Tasks]{\syndy{}: \underline{Syn}thetic \underline{Dy}namic Dataset Generation \\Framework for Misinformation Tasks}

\author{Michael Shliselberg}
\email{michael.shliselberg@uconn.edu}
\affiliation{%
  \institution{
  University of Connecticut}
  \city{Storrs}
  \state{Connecticut}
  \country{USA}
  \postcode{06269}
}

\author{Ashkan Kazemi}
\email{ashkan@meedan.com}
\affiliation{%
  \institution{Meedan}
  \streetaddress{575 Market St., 4th Floor}
  \city{San Francisco}
  \state{California}
  \country{USA}
  \postcode{94105}
}
\orcid{0000-0002-2475-1007}

\author{Scott A. Hale}\authornote{Equally contributing senior corresponding authors.}
\email{scott@meedan.com}
\affiliation{%
  \institution{Meedan \& University of Oxford}
  \streetaddress{1 St Giles}
  \city{Oxford}
  \country{UK}
  \postcode{OX1 3JS}
}
\orcid{0000-0002-6894-4951}

\author{Shiri Dori-Hacohen}\authornotemark[1]
\email{shiridh@uconn.edu}
\affiliation{%
  \institution{
  University of Connecticut}
  \city{Storrs}
  \state{Connecticut}
  \country{USA}
  \postcode{06269}
}

\begin{abstract}
Diaspora communities are disproportionately impacted by off-the-radar misinformation and often neglected by mainstream fact-checking efforts, creating a critical need to scale-up efforts of nascent fact-checking initiatives. 
In this paper we present \textbf{\syndy{}}, a framework for \underline{Syn}thetic \underline{Dy}namic Dataset Generation to leverage the capabilities of the largest frontier Large Language Models (LLMs) to train local, specialized language models. To the best of our knowledge, \syndy{} is the first paper utilizing LLMs to create fine-grained synthetic labels for tasks of direct relevance to misinformation mitigation, namely Claim Matching, Topical Clustering, and Claim Relationship Classification. \syndy{} utilizes LLMs and social media queries to automatically generate distantly-supervised, topically-focused datasets with synthetic labels on these three tasks, providing essential tools to scale up human-led fact-checking at a fraction of the cost of human-annotated data. Training on \syndy{}'s generated labels shows improvement over a standard baseline and is not significantly worse compared to training on human labels (which may be infeasible to acquire). \syndy{} is being integrated into Meedan's chatbot tiplines that are used by over 50 organizations, serve over 230K users annually, and automatically distribute human-written fact-checks via messaging apps such as WhatsApp. \syndy{} will also be integrated into our deployed Co·Insights toolkit, enabling low-resource organizations to launch tiplines for their communities. Finally, we envision \syndy{} enabling additional fact-checking tools such as matching new misinformation claims to high-quality explainers on common misinformation topics.  
\end{abstract}



\makeatletter
\gdef\@copyrightpermission{
 \begin{minipage}{0.3\columnwidth}
  \href{https://creativecommons.org/licenses/by/4.0/}{\includegraphics[width=0.90\textwidth]{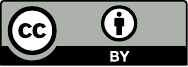}}
 \end{minipage}\hfill
 \begin{minipage}{0.7\columnwidth}
  \href{https://creativecommons.org/licenses/by/4.0/}{This work is licensed under a Creative Commons Attribution International 4.0 License.}
 \end{minipage}
 \vspace{5pt}
}
\makeatother

\maketitle

\vspace{-1mm}
\section{Introduction}

Misinformation predates the Internet \cite{Altay2023}, but combined with the ease of producing and re-sharing content, information on the Internet often spreads without people considering the accuracy of what they see or share \cite{Pennycook2021, Pantazi2021}. 
Misinformation is impactful: it can affect people's beliefs and attitudes \cite{Lewandowsky2017}, how they vote \cite{Vaccari2014}, health choices such as vaccination \cite{Jamieson2021,Pierri2022,Caceres2022}, and (in)action to confront the climate emergency \cite{Benegal2018}. Despite repeated efforts, there are no purely automated systems that successfully and accurately fact-check content~\cite{arnold2020challenges,graves2018understanding}; thus, best practice has shifted from attempts to improve automated fact-checking to building automated tools \textbf{supporting} human fact-checking efforts, such as by matching misinformation claims to existing fact-checks (claim matching)~\cite{kazemi-etal-2021-claim}. 

\begin{figure*}[tb]
     \includegraphics[width=0.95\textwidth]{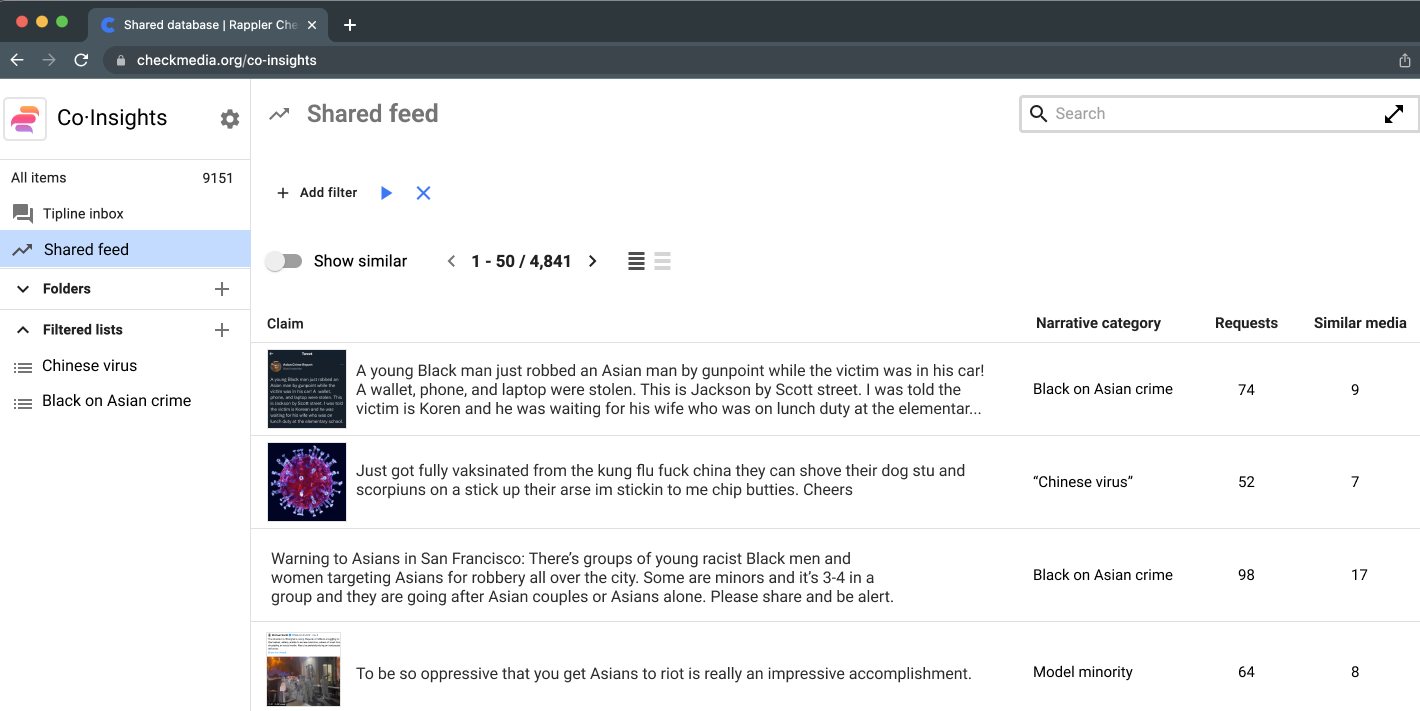}
     \caption{\small{A mockup of the Co·Insights dashboard (built atop Meedan's existing Check platform) with a new ``Narrative Category'' column that will be enabled by \syndy{}'s LLM-generated topic annotations. This, in turn, enables linking to preemptive explainer content in response to end user fact-checking queries on our AAPI nonprofits' dedicated misinformation tiplines.%
     }}
     \label{fig:screenshot}
     \vspace*{-4.5mm}
\end{figure*}

Misinformation targeting diaspora communities is often off the radar of mainstream fact-checking organizations in both host and home countries. For example, the 2017 measles outbreak disproportionately killed more Somali Americans than any other ethnicity group as a result of their Minneapolis community being targeted with anti-vaccine misinformation \cite{leslie2018meales,dyer2017measles}.
While many diaspora communities have formed initiatives to counter misinformation, these community efforts are smaller and lower resourced than mainstream efforts, and there is a critical need for technology to help scale their work. %
Transformer models are useful for a variety of tasks with lower costs and faster throughput than larger frontier models. Nonetheless, it is unreasonable to expect such models to perform well off-the-shelf on the emerging, low resource, and/or out-of-domain topics that are of relevance to diaspora communities. Furthermore, obtaining annotated data may be difficult, costly, time-consuming, or risk vicarious trauma for annotators.

In this demo paper we present \syndy{}, a framework enabling organizations, especially small and/or low-resourced ones, to leverage generative large language models (LLMs) to train local models to perform misinformation-related tasks, reducing the need for costly, time-consuming, and possibly harmful manual annotation. To the best of our knowledge, \syndy{} is the first  paper utilizing LLMs to create fine-grained synthetic labels for tasks relevant to misinformation mitigation, namely \textbf{Claim Matching}, \textbf{Topical Clustering}, and \textbf{Claim Relationship Classification}. As we explain below, each of these tasks is an essential tool to mitigate misinformation by scaling up the impact of human-led fact-checking. Our demonstration video and codebase are both available online.\footnote{Backend: \url{https://github.com/RIET-lab/syndy-service}; UI: \url{https://github.com/RIET-lab/syndy-ui}; Demo Video: \url{https://bit.ly/syndyDemoVideo}}

\section{Team, Project, and Audience}

\syndy{} was developed by the Reducing Information Ecosystem Threats (\riet{}) Lab and Meedan as part of Co·Insights, a large-scale misinformation mitigation initiative funded by the NSF Convergence Accelerator.\footnote{\url{https://meedan.com/project/co-insights}} 
The \rietlab{} at the University of Connecticut focuses on threats to the online ecosystem from an information retrieval lens. Meedan is a \textit{global technology non-profit} that builds software and programmatic initiatives to strengthen journalism, digital literacy, and information accessibility online and off. Meedan develops \textit{open-source tools} for creating and sharing context on digital media through annotation, verification, archival, and translation. %
Led by Meedan, \textbf{Co·Insights} is a unique platform that enables a community-led approach to \textbf{identify}, \textbf{preempt}, and \textbf{respond} to misinformation in minoritized communities. Co·Insights is co-developed with: Asian American \& Pacific Islander (AAPI) organizations operating in the fact-checking space (Piyaoba, Tayo, Viet Fact Check, and DesiFacts)\footnote{\url{https://www.piyaoba.org/} ; \url{https://www.tayohelp.com/} ; \url{https://vietfactcheck.org/} ; \url{https://vietfactcheck.org/} ; \url{https://www.desifacts.org/}} and universities (the Annenberg Public Policy Center (APPC) at the University of Pennsylvania, University of Massachusetts Amherst, University of Connecticut, Rutgers University, and University of Colorado Boulder).

As with Meedan's other misinformation mitigation tools, the \textbf{intended audience for \syndy{}} is fact-checkers, journalists, and community organizations, including especially our Co·Insights partners. Meedan supports over 75 partner organizations, developing technology for newsrooms, fact-checkers, public health professionals, and NGOs. In 2022 alone, Meedan's software served 237K audience members by sending them 200K fact-checks, facilitating 419K conversations with fact-checkers, and matching 43K claims to existing fact-checks~\cite{meedan2022annualreport}. %

\begin{figure*}[htb]
     \includegraphics[width=0.87\textwidth]{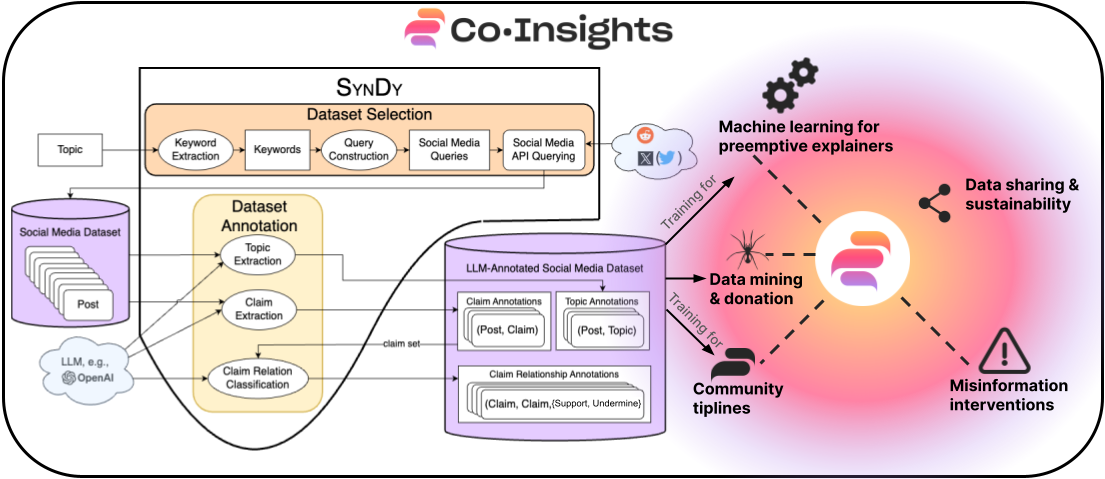}
     \caption{\small{A schematic showing the \syndy{} framework and how it fits within the larger Co·Insights initiative.
     \syndy{}'s output is an LLM-annotated dataset which includes labeled tuples for post-claim, post-topic, and claim-claim-relationships labeled as \inlinemath{\{Support,Undermine\}}. The labels are then utilized as training data to improve misinformation-related tasks. The Dataset Annotation stage can be applied to social media datasets beyond those selected by \syndy{}.}}
     \label{fig:schematic}
     \vspace*{-4.5mm}
\end{figure*}

\vspace{-1mm}
\section{Motivating Challenges} 

\syndy{} arose from \textit{our partners' pain points}, with the goal of better serving them and their communities. We aim to address two real-world challenges: \textit{\textbf{neglected topics}} in mainstream misinformation response efforts and scaling response efforts to \textit{\textbf{narrative themes}}; these challenges demand fine-tuned models for a few \textbf{\textit{misinformation-related tasks}}. 

\textbf{Challenge \#1: Neglected Topics in Mainstream Misinformation Mitigation}. Co·Insights commenced with extensive customer discovery \cite{blank2020startup} with our AAPI organizations. We learned that misinformation content targeting their communities is too often out-of-sight from mainstream response efforts. Such content is in languages, regarding topics, and spreads on platforms beyond those usually considered in the US. \syndy{}'s main objective, therefore, is to improve coverage and performance of fact-checking pipelines, specifically on emerging and/or niche topics where training data is neither readily available nor easy to procure. \syndy{} was also designed with the goal of full integration into Meedan's existing, claim matching pipelines and chatbot tiplines. 

\textbf{Challenge \#2: Scaling Misinformation Efforts to Narrative Themes}. APPC highlighted to us a challenge with current human-led fact-checking efforts: they tend to address claims one at a time after they appear, i.e., in a reactive way. Looking at claims in isolation in this manner results in less effective debunking: by the time interventions occur claims are widespread. %
However, claims do not exist in isolation \cite{ophir2021effects,scales2021covid19}, but rather within a larger framework of recurring narrative and topical themes, and even in fairly predictable patterns~\cite{fact2019report}. By mapping individual claims to these larger ``narrative clusters,'' we can move away from the current reactive model to a new proactive model where context explainers, preemptive knowledge (\textit{``pre-bunking''}~\cite{lewandowsky2021countering}), conversation guides, and media literacy materials can be produced and disseminated before harmful claims spread widely. Since one narrative usually encompasses multiple claims, this approach can greatly increase the relevance and impact of each misinformation intervention. Our AAPI partner organizations have identified the ``green card backlog,'' ``zero-dollar shopping,'' and affirmative action (specifically w/r/t AAPI communities) as topics that could benefit from this approach. Therefore, \syndy{} is also driven by a vision of enabling new misinformation tools to map individual claims to narrative themes (see Figure \ref{fig:screenshot}) and facilitate pre-bunking. %

\textbf{Relevant Misinformation Tasks.} %
We set out to improve upon zero-shot performance for three mis\-in\-for\-ma\-tion-related tasks: \textit{Claim Matching}, \textit{Topical Clustering}, and \textit{Claim Relationship Classification}. First, \textit{\textbf{Claim Matching}} algorithms match new pieces of social media content to existing fact-checks \cite{shaar2020known,kazemi-etal-2021-claim}. %
However, in-the-wild performance of claim matching usually degrades compared to performance on benchmark datasets like CheckThat!~\cite{clef-checkthat:2021:LNCS} due to the continual, ongoing evolution of misinformation requiring the continual retraining of models. 
\textit{\textbf{Topical Clustering}} allows us to match claims to prewritten, high-quality explainers on common topics anticipated by experts as discussed in Challenge 2 above.
Finally, \textit{\textbf{Claim Relationship Classification}} helps human fact-checkers understand and prioritize what misinformation claims to address, based on which claims support or undermine one another. 
Crucially, discerning that such relationships exist does not rely on the need to determine the veracity of claims, but relies only on the connections between claims. Though similar to stance detection, we emphasize connections between claims, rather than the reaction of a subject to a claim, as in most Stance Detection tasks. Taken together, these tasks represent essential tools to mitigate misinformation by \textbf{scaling up the impact of human-led fact-checking}. We also envision them enabling new tools, representing a critical next step for the evolution of fact-checking. 

\vspace{-2mm}
\section{Proposed Framework}

We propose a framework for \textbf{Syn}thetic \textbf{Dy}namic Dataset Generation (\syndy{}) that leverages the recent advances in LLMs~\cite{gpt4,chatgpt} to create distantly-supervised, automatically labeled training datasets (despite these models' well-documented limitations~\cite[see, e.g.,][]{stochastic-parrots}). \syndy{} employs LLMs and leverages prompt engineering in various subtasks, chiefly to extract annotations from social media posts. As a proof of concept, we implemented \syndy{} both with Python scripts that run locally, and with scalable AWS infrastructure. Figure \ref{fig:schematic} show the \syndy{} framework, composed of two main steps: Dataset Selection and Dataset Annotation, in the context of the Co·Insights project's broader misinformation mitigation efforts.

\textbf{Dataset Selection.} We receive a topic as input and produce a set of social media posts on that topic as output, predicated on the availability of some form of query language and social media API (e.g., X/Twitter, Reddit, etc.).\footnote{Alternatively, the dataset could be externally supplied and the selection step skipped.} In our experiments, we used the Reddit API. We perform keyword generation for the topic  to construct social media API queries. While keyword extraction can be performed in a variety of ways, in our experiments we provide the stated topic to the GPT-3.5-Turbo API with a prompt to generate two groups of topical keywords: heavily associated keywords that are largely indicative of the topic and lesser associated keywords that can be associated with the topic but are not mutually exclusive. We then create a collection of queries by sampling one from the former and two from the latter and joining them via AND clauses. 

\textbf{Synthetic Dataset Annotation.} To generate synthetic labels for social media datasets, we take advantage of LLMs' strength in lexical and semantic capabilities, rather than domain knowledge (thus significantly reducing the risks of fabrications~\cite{emsley2023chatgpt,rami2023chatbot}). We constructed specialized prompts through a mixture of following community tips and iterative testing.\footnote{Prompts are omitted for space considerations, but are available in our codebase.} Figure \ref{fig:schematic} shows how we utilize LLM queries to generate labeled tuples for claim extraction, topic clustering, and claim relation classification. The former two tasks receive individual posts as input and provide tuples of the form \inlinemath{(Post, Label)}. For Relation Classification, we need to output source and target claims and their corresponding relationship (support/undermine). This task receives the synthetically-labeled \inlinemath{(Post, Claim)} tuples as input; we then use GPT to generate, for each claim extracted in the previous step, the following two items in the tuple. Finally, a Semantic Clustering step helps tackle the issue of different inputs producing reworded but semantically-identical outputs. We utilize a pretrained sentence transformer (SentenceT5-Large) to map all targets/claims into a dense space and cluster closely packed ones into a single target. The clusters are defined deterministically,where all node pairs with a similarity score above some threshold $\tau$ (which we set at 0.95) are within the same cluster. 

\section{Experimental Setup}

We set out to evaluate whether \syndy{}-labeled datasets could be used as \textbf{training data for misinformation-related tasks}. Specifically, we want to evaluate whether they could (a) improve upon zero-shot settings and (b) be substituted for human annotations as training data for these tasks. 
To that end, we use several preexisting datasets to compare an off-the-shelf baseline and fine-tuned models using either \syndy{}-generated synthetic labels or human-labeled data. We evaluate the results on human-labeled data. %
For \textbf{Claim Matching}, we use the CheckThat!2022 dataset for spotting fact-checked claims in tweets~\cite{clef-checkthat:2022:LNCS}. For \textbf{Topical Clustering}, we use the Cardiff NLP multiple-topic dataset~\cite{CardiffTwitterTC} and the Perspectrum Dataset's topics~\cite{perspectrum}, both of which are provided in the form of topic--tweet pairs. Model training for both these tasks involve fine-tuning SentenceT5-Large~\cite{sentenceT5}, and evaluating with Mean Average Precision (MAP). For \textbf{Claim Relation Classification}, we again use the Perspectrum Dataset in its claim and perspective portion, which provides claim-to-claim relationships with a \inlinemath{\{Support, Undermine\}} label. Here, we use a cross-encoder, encoding both the source and target together. %
Training involves fine-tuning Roberta-Large \cite{roberta}. We use Macro F1 to evaluate the claim relationship, as is common in similar tasks such as stance detection and entailment. We omit further configuration details for space considerations, but our codebase is available.

\section{Results}

\begin{table}
\small
\centering
\resizebox{0.95\columnwidth}{!}{ %
\begin{tabular}{l|ccc}
\hline
 & \multicolumn{3}{c}{\textbf{Train Sets}} \\
\textbf{Test Sets} & Baseline & Human Labels & Synthetic Labels \\ \hline \hline
CheckThat!'22 & .824 & \textbf{.927} & \textbf{.927} \\
CardiffNLP & .660 & \textbf{.753} & .702 \\
Pers.-Clustering & .793 & .800 & \textbf{.828} \\ \hline
Pers.-Classification & -- & \textbf{.907} & .863 \\ \hline
\end{tabular}
}
\caption{\small{Experiment results for claim matching (MAP@20; row 1), topical clustering (MAP@20; rows 2 \& 3), and claim relationship classification (F1 Macro; row 4) tasks. Baseline for matching \& clustering is SentenceT5-Large with no task-specific training.}}
\vspace*{-7mm}
\label{tab:main}
\end{table}
\raggedbottom

Results are presented in Table \ref{tab:main}. Across both the Claim Matching and Topic Clustering tasks, our fine-tuned models utilizing \syndy{} outperformed the Sentence-T5-Large baseline, which was \syndy{}'s explicit goal. In the Claim Matching experiment, our generated and original train sets achieve similar MAP@20, both markedly improving over the pretrained model. In the context of Topical Clustering, the model fine-tuned on synthetic data slightly trails human annotated fine-tuning for the CardiffNLP dataset; perhaps surprisingly, on the Perspectrum-Topics dataset, training on the synthetic dataset actually improved performance above training on human data, possibly due to the higher number of annotations \syndy{} generated.
Finally, for Claim Relation Classification, training on \syndy{}-generated labels yields a Macro-F1 of 0.863, which is lower than the 0.907 from the original dataset, but still promising. 
Across all tasks, models trained on synthetic labels perform no worse than -6.8\% relative to models trained on human annotated datasets, and provide a minimum 4.4\% (on Perspectrum topical clustering) and maximum 12.5\% (on CheckThat!'22 claim matching) relative improvement over a zero-shot baseline across different experimental settings. Overall, models trained on \syndy{}-generated labels demonstrate comparable performance to training on human-annotated data. %

\section{Conclusion}

In this demonstration paper, we presented \syndy{}, a distantly-supervised framework for automatically generating datasets with synthetic labels, with the goal of improving coverage and precision of misinformation related tasks. Specifically, we focused on Claim Matching, Topical Clustering, and Claim Relation Classification. We demonstrated that on these tasks, \syndy{}-generated datasets and synthetic labels can compete with human annotations. %
We leave more direct comparisons of LLM-generated labels to human-annotated labels, the task of narrative theme matching, and cross-lingual challenges to future work. \syndy{} 
supports misinformation response efforts %
in an ever changing information environment. %

\subsubsection*{Acknowledgements}
\small
We thank our collaborators on Co·Insights for inspiring \syndy{} and extensive feedback; Aidan Kierans for LaTeX assistance; and our anonymous reviewers for thoughtful feedback. 
This material is based in part upon work supported by the National Science Foundation Convergence Accelerator (contract 49100421C0035), by the National Science Foundation Program on Fairness in Artificial Intelligence in Collaboration with Amazon (\#IIS-2147305), and by a grant from The Alan Turing Institute. Any opinions, findings and conclusions or recommendations expressed in this material are those of the author(s) and do not necessarily reflect the views of the National Science Foundation or The Alan Turing Institute.%
\balance

\begin{thebibliography}{32}


\ifx \showCODEN    \undefined \def \showCODEN     #1{\unskip}     \fi
\ifx \showDOI      \undefined \def \showDOI       #1{#1}\fi
\ifx \showISBNx    \undefined \def \showISBNx     #1{\unskip}     \fi
\ifx \showISBNxiii \undefined \def \showISBNxiii  #1{\unskip}     \fi
\ifx \showISSN     \undefined \def \showISSN      #1{\unskip}     \fi
\ifx \showLCCN     \undefined \def \showLCCN      #1{\unskip}     \fi
\ifx \shownote     \undefined \def \shownote      #1{#1}          \fi
\ifx \showarticletitle \undefined \def \showarticletitle #1{#1}   \fi
\ifx \showURL      \undefined \def \showURL       {\relax}        \fi
\providecommand\bibfield[2]{#2}
\providecommand\bibinfo[2]{#2}
\providecommand\natexlab[1]{#1}
\providecommand\showeprint[2][]{arXiv:#2}

\bibitem[Altay et~al\mbox{.}(2023)]%
        {Altay2023}
\bibfield{author}{\bibinfo{person}{Sacha Altay}, \bibinfo{person}{Manon Berriche}, {and} \bibinfo{person}{Alberto Acerbi}.} \bibinfo{year}{2023}\natexlab{}.
\newblock \showarticletitle{Misinformation on Misinformation: Conceptual and Methodological Challenges}.
\newblock \bibinfo{journal}{\emph{Social Media + Society}} \bibinfo{volume}{9}, \bibinfo{number}{1} (\bibinfo{year}{2023}).
\newblock
\urldef\tempurl%
\url{https://doi.org/10.1177/20563051221150412}
\showDOI{\tempurl}


\bibitem[Antypas et~al\mbox{.}(2022)]%
        {CardiffTwitterTC}
\bibfield{author}{\bibinfo{person}{Dimosthenis Antypas}, \bibinfo{person}{Asahi Ushio}, \bibinfo{person}{Jos{\'e} Camacho-Collados}, \bibinfo{person}{Leonardo Neves}, \bibinfo{person}{V'itor Silva}, {and} \bibinfo{person}{Francesco Barbieri}.} \bibinfo{year}{2022}\natexlab{}.
\newblock \showarticletitle{Twitter Topic Classification}.
\newblock \bibinfo{journal}{\emph{ArXiv}}  \bibinfo{volume}{abs/2209.09824} (\bibinfo{year}{2022}).
\newblock


\bibitem[Arnold(2020)]%
        {arnold2020challenges}
\bibfield{author}{\bibinfo{person}{Phoebe Arnold}.} \bibinfo{year}{2020}\natexlab{}.
\newblock \showarticletitle{The challenges of online fact checking}.
\newblock \bibinfo{journal}{\emph{Full fact}}  \bibinfo{volume}{17} (\bibinfo{year}{2020}).
\newblock


\bibitem[Benegal and Scruggs(2018)]%
        {Benegal2018}
\bibfield{author}{\bibinfo{person}{Salil~D. Benegal} {and} \bibinfo{person}{Lyle~A. Scruggs}.} \bibinfo{year}{2018}\natexlab{}.
\newblock \showarticletitle{Correcting misinformation about climate change: the impact of partisanship in an experimental setting}.
\newblock \bibinfo{journal}{\emph{Climatic Change}} \bibinfo{volume}{148}, \bibinfo{number}{1} (\bibinfo{year}{2018}), \bibinfo{pages}{61--80}.
\newblock
\urldef\tempurl%
\url{https://doi.org/10.1007/s10584-018-2192-4}
\showURL{%
\tempurl}


\bibitem[Blank and Dorf(2020)]%
        {blank2020startup}
\bibfield{author}{\bibinfo{person}{Steve Blank} {and} \bibinfo{person}{Bob Dorf}.} \bibinfo{year}{2020}\natexlab{}.
\newblock \bibinfo{booktitle}{\emph{The startup owner's manual: The step-by-step guide for building a great company}}.
\newblock \bibinfo{publisher}{John Wiley \& Sons}.
\newblock


\bibitem[Chen et~al\mbox{.}(2019)]%
        {perspectrum}
\bibfield{author}{\bibinfo{person}{Sihao Chen}, \bibinfo{person}{Daniel Khashabi}, \bibinfo{person}{Wenpeng Yin}, \bibinfo{person}{Chris Callison{-}Burch}, {and} \bibinfo{person}{Dan Roth}.} \bibinfo{year}{2019}\natexlab{}.
\newblock \showarticletitle{Seeing Things from a Different Angle: Discovering Diverse Perspectives about Claims}.
\newblock \bibinfo{journal}{\emph{CoRR}}  \bibinfo{volume}{abs/1906.03538} (\bibinfo{year}{2019}).
\newblock
\showeprint[arXiv]{1906.03538}
\urldef\tempurl%
\url{http://arxiv.org/abs/1906.03538}
\showURL{%
\tempurl}


\bibitem[Chiesurin et~al\mbox{.}(2023)]%
        {stochastic-parrots}
\bibfield{author}{\bibinfo{person}{Sabrina Chiesurin}, \bibinfo{person}{Dimitris Dimakopoulos}, \bibinfo{person}{Marco Antonio~Sobrevilla Cabezudo}, \bibinfo{person}{Arash Eshghi}, \bibinfo{person}{Ioannis Papaioannou}, \bibinfo{person}{Verena Rieser}, {and} \bibinfo{person}{Ioannis Konstas}.} \bibinfo{year}{2023}\natexlab{}.
\newblock \bibinfo{title}{The Dangers of trusting Stochastic Parrots: Faithfulness and Trust in Open-domain Conversational Question Answering}.
\newblock
\newblock
\showeprint[arxiv]{2305.16519}~[cs.CL]


\bibitem[Dyer(2017)]%
        {dyer2017measles}
\bibfield{author}{\bibinfo{person}{Owen Dyer}.} \bibinfo{year}{2017}\natexlab{}.
\newblock \showarticletitle{Measles outbreak in Somali American community follows anti-vaccine talks}.
\newblock \bibinfo{journal}{\emph{BMJ}}  \bibinfo{volume}{357} (\bibinfo{year}{2017}).
\newblock
\urldef\tempurl%
\url{https://doi.org/10.1136/bmj.j2378}
\showDOI{\tempurl}
\showeprint{https://www.bmj.com/content/357/bmj.j2378.full.pdf}


\bibitem[Emsley(2023)]%
        {emsley2023chatgpt}
\bibfield{author}{\bibinfo{person}{Robin Emsley}.} \bibinfo{year}{2023}\natexlab{}.
\newblock \showarticletitle{ChatGPT: these are not hallucinations--they’re fabrications and falsifications}.
\newblock \bibinfo{journal}{\emph{Schizophrenia}} \bibinfo{volume}{9}, \bibinfo{number}{1} (\bibinfo{year}{2023}), \bibinfo{pages}{52}.
\newblock


\bibitem[Ferreira~Caceres et~al\mbox{.}(2022)]%
        {Caceres2022}
\bibfield{author}{\bibinfo{person}{Maria~Mercedes Ferreira~Caceres}, \bibinfo{person}{Juan~Pablo Sosa}, \bibinfo{person}{Jannel~A Lawrence}, \bibinfo{person}{Cristina Sestacovschi}, \bibinfo{person}{Atiyah Tidd-Johnson}, \bibinfo{person}{Muhammad Haseeb~Ui Rasool}, \bibinfo{person}{Vinay~Kumar Gadamidi}, \bibinfo{person}{Saleha Ozair}, \bibinfo{person}{Krunal Pandav}, \bibinfo{person}{Claudia Cuevas-Lou}, \bibinfo{person}{Matthew Parrish}, \bibinfo{person}{Ivan Rodriguez}, {and} \bibinfo{person}{Javier~Perez Fernandez}.} \bibinfo{year}{2022}\natexlab{}.
\newblock \showarticletitle{The impact of misinformation on the COVID-19 pandemic}.
\newblock \bibinfo{journal}{\emph{{AIMS} Public Health}} \bibinfo{volume}{9}, \bibinfo{number}{2} (\bibinfo{year}{2022}).
\newblock
\urldef\tempurl%
\url{https://doi.org/10.3934%2Fpublichealth.2022018}
\showURL{%
\tempurl}


\bibitem[{Full Fact}(2019)]%
        {fact2019report}
\bibfield{author}{\bibinfo{person}{{Full Fact}}.} \bibinfo{year}{2019}\natexlab{}.
\newblock \bibinfo{title}{Report on the Facebook Third Party Fact Checking programme}.
\newblock
\newblock


\bibitem[Graves(2018)]%
        {graves2018understanding}
\bibfield{author}{\bibinfo{person}{D Graves}.} \bibinfo{year}{2018}\natexlab{}.
\newblock \showarticletitle{Understanding the promise and limits of automated fact-checking}.
\newblock \bibinfo{journal}{\emph{Reuters Institute for the Study of Journalism}} (\bibinfo{year}{2018}).
\newblock


\bibitem[Hatem et~al\mbox{.}(2023)]%
        {rami2023chatbot}
\bibfield{author}{\bibinfo{person}{Rami Hatem}, \bibinfo{person}{Brianna Simmons}, {and} \bibinfo{person}{Joseph~E. Thornton}.} \bibinfo{year}{2023}\natexlab{}.
\newblock \showarticletitle{{Chatbot Confabulations Are Not Hallucinations}}.
\newblock \bibinfo{journal}{\emph{JAMA Internal Medicine}} \bibinfo{volume}{183}, \bibinfo{number}{10} (\bibinfo{date}{10} \bibinfo{year}{2023}), \bibinfo{pages}{1177--1177}.
\newblock
\showISSN{2168-6106}
\urldef\tempurl%
\url{https://doi.org/10.1001/jamainternmed.2023.4231}
\showDOI{\tempurl}


\bibitem[Jamieson et~al\mbox{.}(2021)]%
        {Jamieson2021}
\bibfield{author}{\bibinfo{person}{Kathleen~Hall Jamieson}, \bibinfo{person}{Daniel Romer}, \bibinfo{person}{Patrick~E. Jamieson}, \bibinfo{person}{Kenneth~M. Winneg}, {and} \bibinfo{person}{Josh Pasek}.} \bibinfo{year}{2021}\natexlab{}.
\newblock \showarticletitle{The role of non–COVID-specific and COVID-specific factors in predicting a shift in willingness to vaccinate: A panel study}.
\newblock \bibinfo{journal}{\emph{Proceedings of the National Academy of Sciences}} \bibinfo{volume}{118}, \bibinfo{number}{52} (\bibinfo{year}{2021}), \bibinfo{pages}{e2112266118}.
\newblock
\urldef\tempurl%
\url{https://doi.org/10.1073/pnas.2112266118}
\showDOI{\tempurl}
\showeprint{https://www.pnas.org/doi/pdf/10.1073/pnas.2112266118}


\bibitem[Kazemi et~al\mbox{.}(2021)]%
        {kazemi-etal-2021-claim}
\bibfield{author}{\bibinfo{person}{Ashkan Kazemi}, \bibinfo{person}{Kiran Garimella}, \bibinfo{person}{Devin Gaffney}, {and} \bibinfo{person}{Scott Hale}.} \bibinfo{year}{2021}\natexlab{}.
\newblock \showarticletitle{Claim Matching Beyond {E}nglish to Scale Global Fact-Checking}. In \bibinfo{booktitle}{\emph{Proceedings of the 59th Annual Meeting of the Association for Computational Linguistics and the 11th International Joint Conference on Natural Language Processing (Volume 1: Long Papers)}}. \bibinfo{publisher}{Association for Computational Linguistics}, \bibinfo{address}{Online}, \bibinfo{pages}{4504--4517}.
\newblock
\urldef\tempurl%
\url{https://doi.org/10.18653/v1/2021.acl-long.347}
\showDOI{\tempurl}


\bibitem[Leslie et~al\mbox{.}(2018)]%
        {leslie2018meales}
\bibfield{author}{\bibinfo{person}{Timothy~F. Leslie}, \bibinfo{person}{Paul~L. Delamater}, {and} \bibinfo{person}{Y.~Tony Yang}.} \bibinfo{year}{2018}\natexlab{}.
\newblock \showarticletitle{It could have been much worse: The {Minnesota} measles outbreak of 2017}.
\newblock \bibinfo{journal}{\emph{Vaccine}} \bibinfo{volume}{36}, \bibinfo{number}{14} (\bibinfo{year}{2018}), \bibinfo{pages}{1808--1810}.
\newblock
\showISSN{0264-410X}
\urldef\tempurl%
\url{https://doi.org/10.1016/j.vaccine.2018.02.086}
\showDOI{\tempurl}


\bibitem[Lewandowsky et~al\mbox{.}(2017)]%
        {Lewandowsky2017}
\bibfield{author}{\bibinfo{person}{Stephan Lewandowsky}, \bibinfo{person}{Ullrich~K.H. Ecker}, {and} \bibinfo{person}{John Cook}.} \bibinfo{year}{2017}\natexlab{}.
\newblock \showarticletitle{Beyond Misinformation: Understanding and Coping with the “Post-Truth” Era}.
\newblock \bibinfo{journal}{\emph{Journal of Applied Research in Memory and Cognition}} \bibinfo{volume}{6}, \bibinfo{number}{4} (\bibinfo{year}{2017}), \bibinfo{pages}{353--369}.
\newblock
\showISSN{2211-3681}
\urldef\tempurl%
\url{https://doi.org/10.1016/j.jarmac.2017.07.008}
\showDOI{\tempurl}


\bibitem[Lewandowsky and Van Der~Linden(2021)]%
        {lewandowsky2021countering}
\bibfield{author}{\bibinfo{person}{Stephan Lewandowsky} {and} \bibinfo{person}{Sander Van Der~Linden}.} \bibinfo{year}{2021}\natexlab{}.
\newblock \showarticletitle{Countering misinformation and fake news through inoculation and prebunking}.
\newblock \bibinfo{journal}{\emph{European Review of Social Psychology}} \bibinfo{volume}{32}, \bibinfo{number}{2} (\bibinfo{year}{2021}), \bibinfo{pages}{348--384}.
\newblock


\bibitem[Liu et~al\mbox{.}(2019)]%
        {roberta}
\bibfield{author}{\bibinfo{person}{Yinhan Liu}, \bibinfo{person}{Myle Ott}, \bibinfo{person}{Naman Goyal}, \bibinfo{person}{Jingfei Du}, \bibinfo{person}{Mandar Joshi}, \bibinfo{person}{Danqi Chen}, \bibinfo{person}{Omer Levy}, \bibinfo{person}{Mike Lewis}, \bibinfo{person}{Luke Zettlemoyer}, {and} \bibinfo{person}{Veselin Stoyanov}.} \bibinfo{year}{2019}\natexlab{}.
\newblock \showarticletitle{RoBERTa: {A} Robustly Optimized {BERT} Pretraining Approach}.
\newblock \bibinfo{journal}{\emph{CoRR}}  \bibinfo{volume}{abs/1907.11692} (\bibinfo{year}{2019}).
\newblock
\showeprint[arXiv]{1907.11692}
\urldef\tempurl%
\url{http://arxiv.org/abs/1907.11692}
\showURL{%
\tempurl}


\bibitem[{Meedan, Inc.}(2022)]%
        {meedan2022annualreport}
\bibfield{author}{\bibinfo{person}{{Meedan, Inc.}}} \bibinfo{year}{2022}\natexlab{}.
\newblock \bibinfo{title}{Annual Report 2022}.
\newblock
\newblock
\urldef\tempurl%
\url{https://meedan.com/post/annual-report-2022}
\showURL{%
\tempurl}


\bibitem[Nakov et~al\mbox{.}(2022)]%
        {clef-checkthat:2022:LNCS}
\bibfield{author}{\bibinfo{person}{Preslav Nakov}, \bibinfo{person}{Alberto Barr\'{o}n-Cede\~{n}o}, \bibinfo{person}{Giovanni Da~San~Martino}, \bibinfo{person}{Firoj Alam}, \bibinfo{person}{Julia~Maria Stru\ss{}}, \bibinfo{person}{Thomas Mandl}, \bibinfo{person}{Rub\'{e}n M\'{\i}guez}, \bibinfo{person}{Tommaso Caselli}, \bibinfo{person}{Mucahid Kutlu}, \bibinfo{person}{Wajdi Zaghouani}, \bibinfo{person}{Chengkai Li}, \bibinfo{person}{Shaden Shaar}, \bibinfo{person}{Gautam~Kishore Shahi}, \bibinfo{person}{Hamdy Mubarak}, \bibinfo{person}{Alex Nikolov}, \bibinfo{person}{Nikolay Babulkov}, \bibinfo{person}{Yavuz~Selim Kartal}, \bibinfo{person}{Javier Beltr\'{a}n}, \bibinfo{person}{Michael Wiegand}, \bibinfo{person}{Melanie Siegel}, {and} \bibinfo{person}{Juliane K{\"o}hler}.} \bibinfo{year}{2022}\natexlab{}.
\newblock \showarticletitle{Overview of the {CLEF}-2022 {CheckThat}! Lab on Fighting the {COVID-19} Infodemic and Fake News Detection}. In \bibinfo{booktitle}{\emph{Proceedings of the 13th International Conference of the CLEF Association: Information Access Evaluation meets Multilinguality, Multimodality, and Visualization}} \emph{(\bibinfo{series}{CLEF~'2022})}. \bibinfo{address}{Bologna, Italy}.
\newblock


\bibitem[Nakov et~al\mbox{.}(2021)]%
        {clef-checkthat:2021:LNCS}
\bibfield{author}{\bibinfo{person}{Preslav Nakov}, \bibinfo{person}{Giovanni Da~San Martino}, \bibinfo{person}{Tamer Elsayed}, \bibinfo{person}{Alberto Barr{\'{o}}n{-}Cede{\~{n}}o}, \bibinfo{person}{Rub{\'{e}}n M{\'{\i}}guez}, \bibinfo{person}{Shaden Shaar}, \bibinfo{person}{Firoj Alam}, \bibinfo{person}{Fatima Haouari}, \bibinfo{person}{Maram Hasanain}, \bibinfo{person}{Watheq Mansour}, \bibinfo{person}{Bayan Hamdan}, \bibinfo{person}{Zien~Sheikh Ali}, \bibinfo{person}{Nikolay Babulkov}, \bibinfo{person}{Alex Nikolov}, \bibinfo{person}{Gautam~Kishore Shahi}, \bibinfo{person}{Julia~Maria Stru{\ss}}, \bibinfo{person}{Thomas Mandl}, \bibinfo{person}{M{\"{u}}cahid Kutlu}, {and} \bibinfo{person}{Yavuz~Selim Kartal}.} \bibinfo{year}{2021}\natexlab{}.
\newblock \showarticletitle{Overview of the {CLEF-2021} CheckThat! Lab on Detecting Check-Worthy Claims, Previously Fact-Checked Claims, and Fake News}.
\newblock \bibinfo{journal}{\emph{CoRR}}  \bibinfo{volume}{abs/2109.12987} (\bibinfo{year}{2021}).
\newblock
\showeprint[arXiv]{2109.12987}
\urldef\tempurl%
\url{https://arxiv.org/abs/2109.12987}
\showURL{%
\tempurl}


\bibitem[Ni et~al\mbox{.}(2021)]%
        {sentenceT5}
\bibfield{author}{\bibinfo{person}{Jianmo Ni}, \bibinfo{person}{Gustavo~Hern{\'{a}}ndez {\'{A}}brego}, \bibinfo{person}{Noah Constant}, \bibinfo{person}{Ji Ma}, \bibinfo{person}{Keith~B. Hall}, \bibinfo{person}{Daniel Cer}, {and} \bibinfo{person}{Yinfei Yang}.} \bibinfo{year}{2021}\natexlab{}.
\newblock \showarticletitle{Sentence-T5: Scalable Sentence Encoders from Pre-trained Text-to-Text Models}.
\newblock \bibinfo{journal}{\emph{CoRR}}  \bibinfo{volume}{abs/2108.08877} (\bibinfo{year}{2021}).
\newblock
\showeprint[arXiv]{2108.08877}
\urldef\tempurl%
\url{https://arxiv.org/abs/2108.08877}
\showURL{%
\tempurl}


\bibitem[OpenAI(2023)]%
        {gpt4}
\bibfield{author}{\bibinfo{person}{OpenAI}.} \bibinfo{year}{2023}\natexlab{}.
\newblock \bibinfo{title}{GPT-4 Technical Report}.
\newblock
\newblock
\showeprint[arxiv]{2303.08774}~[cs.CL]


\bibitem[Ophir and Jamieson(2021)]%
        {ophir2021effects}
\bibfield{author}{\bibinfo{person}{Yotam Ophir} {and} \bibinfo{person}{Kathleen~Hall Jamieson}.} \bibinfo{year}{2021}\natexlab{}.
\newblock \showarticletitle{The effects of media narratives about failures and discoveries in science on beliefs about and support for science}.
\newblock \bibinfo{journal}{\emph{Public Understanding of Science}} \bibinfo{volume}{30}, \bibinfo{number}{8} (\bibinfo{year}{2021}), \bibinfo{pages}{1008--1023}.
\newblock
\urldef\tempurl%
\url{https://doi.org/10.1177/09636625211012630}
\showDOI{\tempurl}
\newblock
\shownote{PMID: 34000907}.


\bibitem[Ouyang et~al\mbox{.}(2022)]%
        {chatgpt}
\bibfield{author}{\bibinfo{person}{Long Ouyang}, \bibinfo{person}{Jeff Wu}, \bibinfo{person}{Xu Jiang}, \bibinfo{person}{Diogo Almeida}, \bibinfo{person}{Carroll~L. Wainwright}, \bibinfo{person}{Pamela Mishkin}, \bibinfo{person}{Chong Zhang}, \bibinfo{person}{Sandhini Agarwal}, \bibinfo{person}{Katarina Slama}, \bibinfo{person}{Alex Ray}, \bibinfo{person}{John Schulman}, \bibinfo{person}{Jacob Hilton}, \bibinfo{person}{Fraser Kelton}, \bibinfo{person}{Luke Miller}, \bibinfo{person}{Maddie Simens}, \bibinfo{person}{Amanda Askell}, \bibinfo{person}{Peter Welinder}, \bibinfo{person}{Paul Christiano}, \bibinfo{person}{Jan Leike}, {and} \bibinfo{person}{Ryan Lowe}.} \bibinfo{year}{2022}\natexlab{}.
\newblock \bibinfo{title}{Training language models to follow instructions with human feedback}.
\newblock
\newblock
\showeprint[arxiv]{2203.02155}~[cs.CL]


\bibitem[Pantazi et~al\mbox{.}(2021)]%
        {Pantazi2021}
\bibfield{author}{\bibinfo{person}{Myrto Pantazi}, \bibinfo{person}{Scott Hale}, {and} \bibinfo{person}{Olivier Klein}.} \bibinfo{year}{2021}\natexlab{}.
\newblock \showarticletitle{Social and Cognitive Aspects of the Vulnerability to Political Misinformation}.
\newblock \bibinfo{journal}{\emph{Political Psychology}} \bibinfo{volume}{42}, \bibinfo{number}{S1} (\bibinfo{date}{Dec.} \bibinfo{year}{2021}), \bibinfo{pages}{267–304}.
\newblock
\showISSN{1467-9221}
\urldef\tempurl%
\url{https://doi.org/10.1111/pops.12797}
\showDOI{\tempurl}


\bibitem[Pennycook and Rand(2021)]%
        {Pennycook2021}
\bibfield{author}{\bibinfo{person}{Gordon Pennycook} {and} \bibinfo{person}{David~G. Rand}.} \bibinfo{year}{2021}\natexlab{}.
\newblock \showarticletitle{The Psychology of Fake News}.
\newblock \bibinfo{journal}{\emph{Trends in Cognitive Sciences}} \bibinfo{volume}{25}, \bibinfo{number}{5} (\bibinfo{date}{May} \bibinfo{year}{2021}), \bibinfo{pages}{388–402}.
\newblock
\showISSN{1364-6613}
\urldef\tempurl%
\url{https://doi.org/10.1016/j.tics.2021.02.007}
\showDOI{\tempurl}


\bibitem[Pierri et~al\mbox{.}(2022)]%
        {Pierri2022}
\bibfield{author}{\bibinfo{person}{Francesco Pierri}, \bibinfo{person}{Brea~L. Perry}, \bibinfo{person}{Matthew~R. DeVerna}, \bibinfo{person}{Kai-Cheng Yang}, \bibinfo{person}{Alessandro Flammini}, \bibinfo{person}{Filippo Menczer}, {and} \bibinfo{person}{John Bryden}.} \bibinfo{year}{2022}\natexlab{}.
\newblock \showarticletitle{Online misinformation is linked to early {COVID-19} vaccination hesitancy and refusal}.
\newblock \bibinfo{journal}{\emph{Scientific Reports}} \bibinfo{volume}{12}, \bibinfo{number}{1} (\bibinfo{year}{2022}).
\newblock
\urldef\tempurl%
\url{https://doi.org/10.1038/s41598-022-10070-w}
\showURL{%
\tempurl}


\bibitem[Scales et~al\mbox{.}(2021)]%
        {scales2021covid19}
\bibfield{author}{\bibinfo{person}{David Scales}, \bibinfo{person}{Jack Gorman}, {and} \bibinfo{person}{Kathleen~H. Jamieson}.} \bibinfo{year}{2021}\natexlab{}.
\newblock \showarticletitle{The Covid-19 Infodemic — Applying the Epidemiologic Model to Counter Misinformation}.
\newblock \bibinfo{journal}{\emph{New England Journal of Medicine}} (\bibinfo{year}{2021}).
\newblock
\urldef\tempurl%
\url{https://doi.org/10.1056/NEJMp2103798}
\showDOI{\tempurl}


\bibitem[Shaar et~al\mbox{.}(2020)]%
        {shaar2020known}
\bibfield{author}{\bibinfo{person}{Shaden Shaar}, \bibinfo{person}{Nikolay Babulkov}, \bibinfo{person}{Giovanni Da~San~Martino}, {and} \bibinfo{person}{Preslav Nakov}.} \bibinfo{year}{2020}\natexlab{}.
\newblock \showarticletitle{That is a Known Lie: Detecting Previously Fact-Checked Claims}. In \bibinfo{booktitle}{\emph{Proceedings of the 58th Annual Meeting of the Association for Computational Linguistics}}. \bibinfo{publisher}{Association for Computational Linguistics}, \bibinfo{address}{Online}, \bibinfo{pages}{3607--3618}.
\newblock
\urldef\tempurl%
\url{https://doi.org/10.18653/v1/2020.acl-main.332}
\showDOI{\tempurl}


\bibitem[Vaccari and Morini(2014)]%
        {Vaccari2014}
\bibfield{author}{\bibinfo{person}{Cristian Vaccari} {and} \bibinfo{person}{Marco Morini}.} \bibinfo{year}{2014}\natexlab{}.
\newblock \showarticletitle{The Power of Smears in Two American Presidential Campaigns}.
\newblock \bibinfo{journal}{\emph{Journal of Political Marketing}} \bibinfo{volume}{13}, \bibinfo{number}{1-2} (\bibinfo{year}{2014}), \bibinfo{pages}{19--45}.
\newblock
\urldef\tempurl%
\url{https://doi.org/10.1080/15377857.2014.866021}
\showDOI{\tempurl}


\end{thebibliography}

\appendix

\end{document}